\documentclass[10pt,letterpaper]{article}
\usepackage[top=0.85in,marginparwidth=2in]{geometry}
\usepackage[utf8]{inputenc}
\usepackage{cite}
\usepackage{microtype}
\DisableLigatures[f]{encoding = *, family = * }
\usepackage{changepage}
\usepackage[aboveskip=1pt,labelfont=bf,labelsep=period,singlelinecheck=off]{caption}
\usepackage{url}

\usepackage{amsmath}
\usepackage{multirow}
\usepackage{booktabs}
\usepackage{float}
\usepackage{adjustbox}

\makeatletter
\renewcommand{\@biblabel}[1]{\quad#1.}
\makeatother

\usepackage{lastpage,fancyhdr,graphicx}
\usepackage{epstopdf}
\fancyheadoffset[L]{2.25in}
\fancyfootoffset[L]{2.25in}

\usepackage{color}

\definecolor{Gray}{gray}{.25}

\usepackage{graphicx}

\usepackage{sidecap}

\usepackage{wrapfig}
\usepackage[pscoord]{eso-pic}
\usepackage[fulladjust]{marginnote}
\reversemarginpar

\begin{document}
% title goes here:
\begin{flushleft}
{\Large
\textbf\newline{ Distributed focus and digital zoom}
}
\newline
% authors go here:
\\
 Wubin Pang and David J. Brady
\\
\bigskip
Department of Electrical and Computer Engineering, Duke University, Durham, NC 27708, USA

\bigskip
* dbrady@duke.edu

\end{flushleft}
%\title{Distributed focus and digital zoom}

%\author{WUBIN PANG AND DAVID J. BRADY{*}}

%\address{Department of Electrical and Computer Engineering, Duke University, Durham, NC 27708, USA}

%\email{\authormark{*}david.brady@duke.edu}


\section*{Abstract}
 We explore integrated microcamera focus systems for array cameras. We propose a new model for system camera integration relying on fast action focus mechanisms with $<10 mm$ aperture. Rather than reducing resolution or expanding aperture size, such systems can be used in arrays to enable digital zoom. We show that a common mechanism supports camera modules with focal lengths ranging from 25 to 60 mm. Designs for each focal length include a fixed objective lens group and an adjustable back focus group. Increasing the focal power of the front focal group enables the travel range of available microcamera modules to accommodate long focal length systems. We present design examples both discrete and multiscale array camera systems. 

% now start line numbers

%\ocis{(000.0000) General; (000.2700) General science.} % REPLACE WITH CORRECT OCIS CODES FOR YOUR ARTICLE, %MINIMUM OF TWO; Avoid using the OCIS codes for “General” or “General science” whenever possible.
%For a complete list of OCIS codes, visit: https://www.osapublishing.org/oe/submit/ocis/

%%%%%%%%%%%%%%%%%%%%%%%%%%  body  %%%%%%%%%%%%%%%%%%%%%%%%%%
\section*{Introduction}
\label{Intro}
The interchangeable lens camera has been the dominant model of high performance imaging for the past 100 years~\cite{KANG2017376}. However, this model assumes a rough equivalence in the relative value of lenses and camera backs and also, since tolerances for mechanical mounting are not high, is onlyt suitable for relatively large $f/\#$. A new approach, based on integrated camera modules, has emerged over the past 20 years to support cameras in mobile devices. In these systems the camera sensor and the lens are permanently and precisely integrated such that $f/1-f/3$ performance may be obtained consistent with pixel pitch ranging from 1.2 to 2 microns. These integrated modules more accurately reflect the development state of modern sensors, wherein sensor cost is less than \$1. In contrast, manufacturing technologies for macroscopic interchangeable lenses have not been changed substantially over the past many years and the cost of such lenses remains in the \$100 to \$1000 range. The designs presented here suggest that a novel approach to system camera design based on interchangable arrays of integrated microcameras may have advantages with respect to interchangable lens systems, especially with respect to size,  cost and focal accommodation speed. 

As diverse camera systems have appeared, the concept of "35 mm equivalent focal length" has been used to express relationships between the imaging characteristics of optical systems with various sensor formats \cite{cipa}. While equivalent focal length definition simply  matches the field of view of compact camera modules with corresponding 35 mm systems, since the $f/\#$ and pixel size can be much smaller for compact imaging systems the equivalent focal length often expresses a true relationship between actual instantaneous field of view (ifov) of the matched systems. For typical mobile microcameras, the 35 mm equivalent focal length may be 6-8 times greater than the actual focal length. Current mobile devices typically have focal lengths of 4-5 mm, corresponding to 20-30 mm 35 mm equivalent focal length, but if lenses at equivalent $f/\#$ are constructed with focal lengths of 20-60 mm on the same sensor, imaging performance comparable to 100-300 mm focal length interchangeable lenses may be expected. 

Despite rapid advances in computational imaging, physical focus remains essential to high resolution imaging. Since basic quantum mechanical constraints preclude direct measurement of the field on the aperture~\cite{brady2018parallel}, well focused lenses remain the best way to process and capture image data. Many low-cost cameras rely on prime lenses focused at or near the hyperfocal distance, but for focal lengths longer than 5-10 millimeters the hyperfocal distance is unacceptably large and dynamic focus must be used. Longer focal length cameras therefore typically include focus actuators. Such cameras also use mechanical actuators for optical zoom. The most common zoom and focus mechanisms move a group of lens elements in the axial direction. 

Zoom and focus mechanisms present several challenges to the camera designer. First, mechanical focus and zoom mechanisms tend to be complex and expensive. It usually contains many components, a stepper or an ultrasonic motor with large size and high power consumption, a reduction gear and a cam mechanism, a control mechanism and sometimes an additional optical path. Second, the response rate of the focus and zoom mechanism is often slow due to large components with inertial momentum. Finally, since objects at different depths and different field points are generally present, there is no correct solution for focus and zoom settings. Focusing or zooming on one object in the field tends to blur other objects or leave them out entirely.  

The challenge of physical focus has been resolved particularly successfully in support of mobile imaging systems
\cite{gutierrez2007auto}. Voice coil motor (VCM) focus mechanisms are manufactured in volumes of >1B units/year, vastly exceeding the total number of focus mechanisms manufactured for interchangable lens cameras over the past 100 years. These focus mechanisms are extremely inexpensive and reliable. The low cost technologies developed in miniature camera community should not remain confined to miniature camera industry. The systems can be adapted and transferred into the professional camera industry. This paper describes designs using the miniature focusing mechanism (MFM) found in today's cellular phone and camera array architecture to build inexpensive, compact and fast focusing and digital zoom functionality in high resolution camera systems. 

Section 2 of this paper addresses technical issues unique to such designs. First, we seek to elucidate the design principles of lenses with a tightly constrained space at the back end resulted from a miniaturized focusing group. The stop position and the placement of a light bending lens group are two key measures for achieving a well performing and balanced design. Second, we address the limited focal accomodation motion range available in MFM modules. We derive an analytical equation describing the relationship between the driving range and the power of the fixed lens group. This offers a mechanism to match the available driving range to the requirements of long focal length systems.  Then we propose universal focusing modules for reducing manufacturing cost even further. Section 3 then presents concrete design examples using MFM modules. Section 4 considers strategies for using these modules in array camera systems. Where interchangeable lens cameras use lens selection to balance field of view and angular resolution, we propose that future system cameras may use microcamera selection to achieve these targets in array solutions. Multiscale camera designs \cite{brady2012multiscale} in particular offer the potential gigapixel scale compact imaging solutions when integrated with MFM modules. A multiscale design example is presented in section 4.2. 

\section*{Focusing with miniaturized focusing mechanism}
\label{FocusingWithMFM}

\subsection*{Motivation in adopting MFM in non-miniaturized camera}
\label{Motive}

A proper focusing mechanism is critical for creating images comprehensible to human visual understanding. Over the long history of imagery, whether for still photography or videography, capturing subjects with sharp contrast is a required element for a successful imaging system. It is rare for camera captures scene or object only with fixed distance and field depth. Therefore, focusing must be provided for accommodating different object ranges. 

Despite novel emerging focus technologies, such as modifying lens elements with liquid materials or liquid crystals, eliminating the need for focus using computational extended depth of field (DoF)\cite{berge2000variable,dowski1995extended,llull2015image}, axial translation-based focus focusing still dominates the market. In these systems, an actuator drives a group of lens elements back and forth.

In applications with near points in excess of 2 meters and ifov greater than 300 microradians, fixed focus lenses with focal lengths less than 5 mm are satifactory. However, when closer near points and greater resolution is needed, active focus is required. For example, early phone cameras captured VGA-scale resolution with no need for active focus. As pixel count reached and exceeded 2 megapixels, active focus became imperative. As the total resolution comes even higher, more and more sophisticated focusing mechanisms have been necessary. Consequently, total accessible pixel count is one of the critical indicators of the performance of a focusing mechanism. The total pixel count should be calculated as summation of pixel count over every focusing position. Currently, miniaturized camera modules of various kinds are capable of capturing pixel volumes in excess of 10 megapixels. This pixel count can be maintained over 10-20 focal positions. The total accesible pixle range, in excess of 100 megapixels, is comparable to high end digital single lens reflex (DSLRs).
 
 The extraordinarily high information throughput is derived from the superior quality of the miniature optics over traditional ones. Aperture size of small scale tends to have much higher information efficiency than that of large aperture ones due to the scale law of lens design \cite{lohmann1989scaling}. To put it another way, information efficiency indicated as the ratio of number of resolvable spots to that from diffraction limit decreases rapidly as the size of lens grows as a result of proportionally deteriorating geometric aberration. In addition, sophisticated aspheric surfaces are readily available in small aperture optics but very difficult to manufacture in macroscopic lenses. With efficient techniques like injection molding, high quality miniaturized lens can be mass produced under extremely low cost.   
 
 Unlike driving motors in the traditional cameras, such as step motor (STM) and ultrasonic motor (USM), actuators in miniaturized camera are small in size, light in weight and efficient in energy consumption. For example, miniature systems predominately  voice coil motors (VCM) to enable fast, precise, robust and inexpensive focusing. Using small inertial mass and fast electromechanical response, VCMs allow camera to lock on and keep track of high dynamic scenes. Larger size lens naturally has larger mass, which slows the motion.  VCM focus adjustment times are typically less than $10ms$ \cite{murphy2006lens,liu2009design,liu2011experimental}, in constrast with $>0.1s$ for typical full aperture cameras \cite{autofocusspeed,phoneArena}. 
Lastly, the growth of geometric aberration with scale, the complexity of traditional monolithic lens increases non-linearly with information capacity . Previously, we have proposed parallel camera as a straightforward solution to high resolution wide field of view camera development \cite{brady2018parallel}. However,  parallel cameras require independent focusing mechanisms for each individual channel, since each channel points to different direction. A large number of regular sized focusing mechanisms would account for a large amount of cost, space tension and operation overhead. MFMs  provide a timely solution which matches perfectly with the many motives behind parallel camera architecture.  

\subsection*{Focusing with low cost mechanism transferred from miniature cameras}
\label{MiniatureFocus}

The focusing performance of miniaturized cameras has improved rapidly in terms of speed, preciseness, mechanism size and  power consumption over the past decades\cite{liu2008miniaturized,jang2015sensor,hsieh2018design}. Currently available systems also support additional functionalities such as image stabilization (IS) and inexpensive image processing.
However, MFMs operate under severe  space and operating power constraints. To implement it in regular camera, the focusing optics must share same characteristic of miniaturized lens in terms of size, weight and power. 

\begin{figure}[ht!]
\centering\includegraphics[width=7cm]{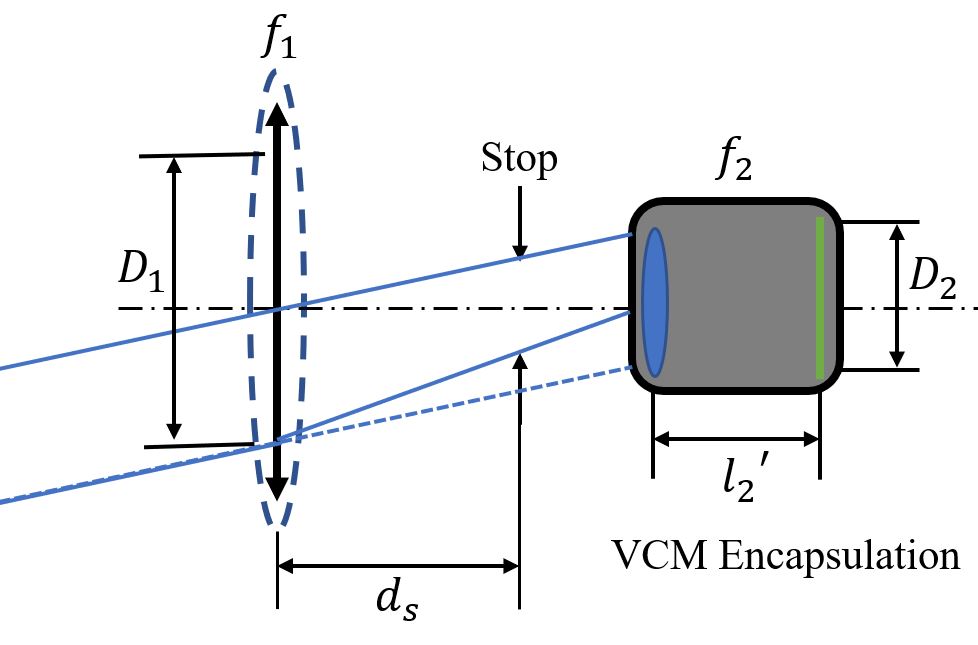}
\caption{Critical parameters in applying MFM into conventional lens architecture.}
\label{fig1}
\end{figure}

To start the design work, we need to specify overall focal length $f$, system aperture $F/\#$ and field of view $FoV$ as usual. A simplified first order model helps clarify the trade-off and constraints associated with such a system. As depicted in Fig.\ref{fig1}, our model contains two optical groups, a front group with focal length $f_1$ and a focusing group in the back with focal length $f_2$. The separation in between is $d$. To complete the model, stop position is another critical parameter to be chosen with care. Here stop position is described as $d_s$ which represents the distance between the stop surface and the front lens group. A well selected combination of these four parameters would help accomplish the design specifications under the permission of the constraints.  
 
The followings are equations expressing three most critical parameters when designing such a system, all parameters are labeled in Fig.\ref{fig1}

\begin{equation}
D_1=2tan(\frac{FoV}{2})\frac{f_1d_s}{f_1-d_s}+\frac{f}{F/\#}
\label{Fir_Eqn}
\end{equation}

\begin{equation}
D_2=2tan(\frac{FoV}{2})\frac{f_1(d-d_s)}{f_1-d_s}+\frac{{l_2}^\prime}{F/\#}
\label{Sec_Eqn}
\end{equation}

\begin{equation}
{l_2}^\prime=\frac{(f_1-d)f_2}{f_1+f_2-d}
\label{Tir_Eqn}
\end{equation}

Where $D_1$ and $D_2$ denote the clear aperture diameter of the front group and the back group respectively, ${l_2}^\prime$ represents back working distance of the back group. $d_s$ in equations is the distance between the front group and the stop surface. On one hand, the employment of MFM imposes two unique constraints on this design work. The aperture diameter $D_2$ and back working distance ${l_2}^\prime$ of the focusing lens group should be small enough in order to fitting into the package of the MFM, whose typical size is less than 6mm in all dimensions. On the other hand, to ensure an acceptable imaging quality which requires well corrected aberration and a relative simple lens build, a good rule of thumb is the aperture diameter of a lens group usually had better not be larger than its focal length, or to express it mathematically, as $D_1\leq |f_1|, D_2\leq |f_2|$.

These parameters impose tight limitations on achievable system $FoV, F/\#$ as well as $f$. Indicated by Eqn.(\ref{Sec_Eqn}), On one hand, by placing the stop close to the focusing group can reduce the clear aperture pressure imposed by the system $FoV$. On the other hand, this manipulation would transfer the aperture pressure towards the front group as manifested by Eqn.(\ref{Fir_Eqn}). Therefore, the choice on positioning of the stop surface $d_s$ can be used to strike a balance between clear apertures of the front group and the back group. 

Small $D_2$ also indicates a relative narrow system $FoV$ compared with regular camera lenses as implied by Eqn.(\ref{Sec_Eqn}). Therefore, small format image sensors should be employed. Moreover,VCM mechanisms are tyipcally designed to match such sensors. Though the noise and low dynamic range performance of small pixel sensors does not match larger systems, the performance has improved significantly due to the massive market demand. Where greater dyanmic range and noise performance is required, it is likely better to use arrays of small pixel sensors or dyanamic sampling using such sensors than to resort to traditional large pixel systems. The rapidly advancing semiconductor industry has driven the price of sensor chip mass produced sensors to extremely low levels, making such array systems less expensive thant larger sensors. Technologies such as back side illumination (BSI) with high photon efficiency also  improves their low light performance and suppresses the noise associated with small pixel pitch. Innovations such as non-planar technology and new materials have made the chips even compact. 

\begin{figure}[ht!]
\centering\includegraphics[width=7cm]{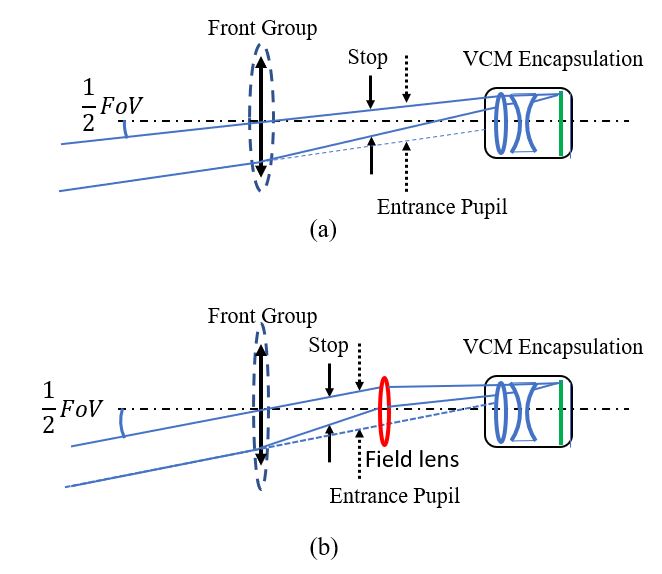}
\caption{FoV of MFM focusing lens. (a) Much limited FoV without using field lens.(b) Enlarged FoV with use of field lens.}
\label{fig2}
\end{figure}

Even with the approach of adjusting the position of stop surface and the employment of small format image sensors, achieving desired $FoV$ still remains a hard task. As demonstrated in Fig.\ref{fig2}, one simple solution is by placing a field lens after the stop surface. This field lens helps by bending light beam from high field angle into the small aperture of the back group.

Besides the tightly constrained space, another challenge associated with the employment of MFM is that a typical VCM can only provide a traveling range greater than $300\mu m$, this is hardly enough for focusing regular lenses. We address this issue next.

\subsection*{Range of travel reduction with increased focal power in front lens group}
\label{TravelingRangeReduction}

Except for rare cases where the whole optical assembly is moved to focus, lens systems focus by moving only one element or a small group of elements. The benefits are manifold. First, moving only a few elements requires less driving force. Second, the optomechanics tends to be less complex and less expensive. Lastly, driving only partial elements offers the opportunity to balance the design in tradeoffs between the characteristics of each group.  Motion parameters  include travel range, travel position resolution and stability. This section explores the reduction of necessary traveling range by means of adjusting lens powers designated on the two lens groups. 
 Image position is related to  object position by the Gauss lens formula $\frac{1}{l^\prime}-\frac{1}{l}=\frac{1}{f}$. Changes in  object range result in changes in the image distance. In practice, designers rarely move the focal plane because it is rather sensitive to misalignment. Lens elements are a better choice for focal adjustment.  Our derivation uses conjugate distance between the object and the image plane denoted by capital $L$ as an independent variable as shown in Eqn.(\ref{Fou_Eqn}). This equation is only a change of form of the Gauss lens formula for emphasizing the relationship between the image distance denoted as $l^\prime$ and the conjugate distance $L$ which can be calculated as $L=l^\prime-l$, $l$ is the object distance and $f$ represents the focal length. By convention, a real object lies to the left of the lens and objects to the right are considered virtual. Distance to the left side of the lens is negative and that to the right is positive. For an image on the right side of its object, we define the conjugate distance $L$ to be positive, otherwise to be negative. The focal length $f$ is positive if the lens has positive power while it is negative for the lens has negative power.  

\begin{equation}
l^\prime=
\begin{cases}
\frac{1}{2}(L-\sqrt{L^2-4Lf}) & \textnormal{if} \quad l\in [-\infty, -2f]\cup(-f,0] \&\& f>0\\
\\
\frac{1}{2}(L+\sqrt{L^2-4Lf}) & \textnormal{if} \quad l\in (-2f,-f]\cup(0, +\infty]\&\& f>0\\
\\
\frac{1}{2}(L-\sqrt{L^2-4Lf}) & \textnormal{if} \quad l\in [-\infty, 0)\cup(-f,-2f] \&\& f<0\\
\\
\frac{1}{2}(L+\sqrt{L^2-4Lf}) & \textnormal{if} \quad l\in (0,-f]\cup(-2f, +\infty]\&\& f<0
\end{cases}
\label{Fou_Eqn}
\end{equation}

All possible imaging cases are accounted for in Eqn.(\ref{Fou_Eqn}). Since the object and image are interchangeable, for a pair of given conjugate distance $L$ and focal length $f$, there are two possible solutions for the image distance $l^\prime$. Imaging with positive lenses is expressed in the first two equations and with negative lenses is in the last two equations. Although the cases are many, we won't dive into each of them in this manuscript. Only these which suit our specific design constraints will be studied and presented.

Among many important parameters for the focusing motion, traveling range determines the depth range of the scene over which the lens system is capable of covering. Generally, a deep depth requirement dictates a long traveling range and a shallow depth for a short traveling range. 
 
\begin{figure}[ht!]
\centering\includegraphics[width=8cm]{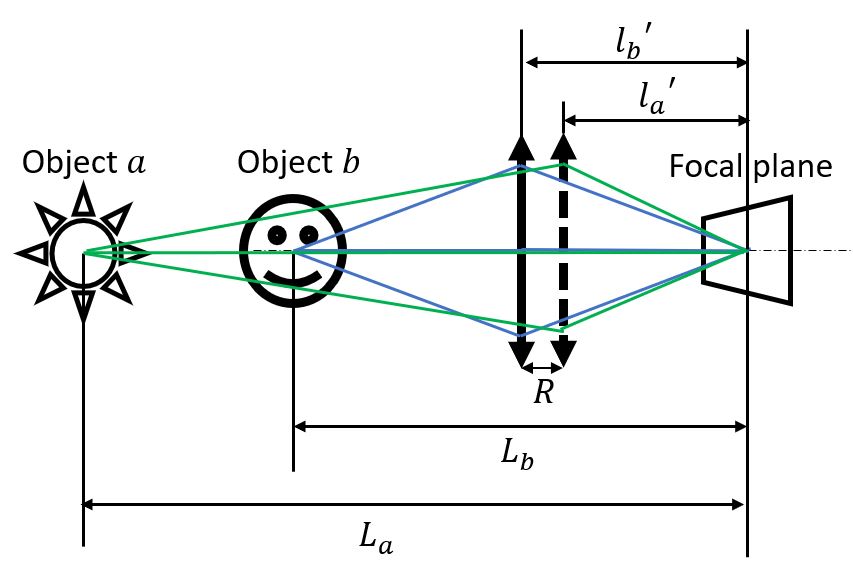}
\caption{Focusing with lens translation in axial direction.}
\label{fig3}
\end{figure}

As illustrated in Fig.\ref{fig3}, assuming a camera changes its focusing point from the object a to object b with conjugate distances of $L_a$ to $L_b$ respectively. After focusing, the image distance changes from ${l_a}^\prime$ to ${l_b}^\prime$. The required traveling range $R$ is given by 

\begin{equation}
R=|\Delta l^\prime|=|{l_a}^\prime-{l_b}^\prime|
\label{Fif_Eqn}
\end{equation}

Here we confine ourselves in cases of photographic or telescopic lenses, in which the object distance is much greater than the focal length of the optics. Therefore, it is safe to assume that $L\gg 4|f|$. Applying this assumption in the first expression in Eqn.(\ref{Fou_Eqn}) and then feeding the result into Eqn.(\ref{Fif_Eqn}), the traveling range $R_o$ can be expressed as

\begin{equation}
R_o=f^2|\frac{1}{L_a}-\frac{1}{L_b}|\approx f^2|\frac{1}{l_a}-\frac{1}{l_b}| 
\label{Six_Eqn}
\end{equation}

\begin{figure}[ht!]
\centering\includegraphics[width=12cm]{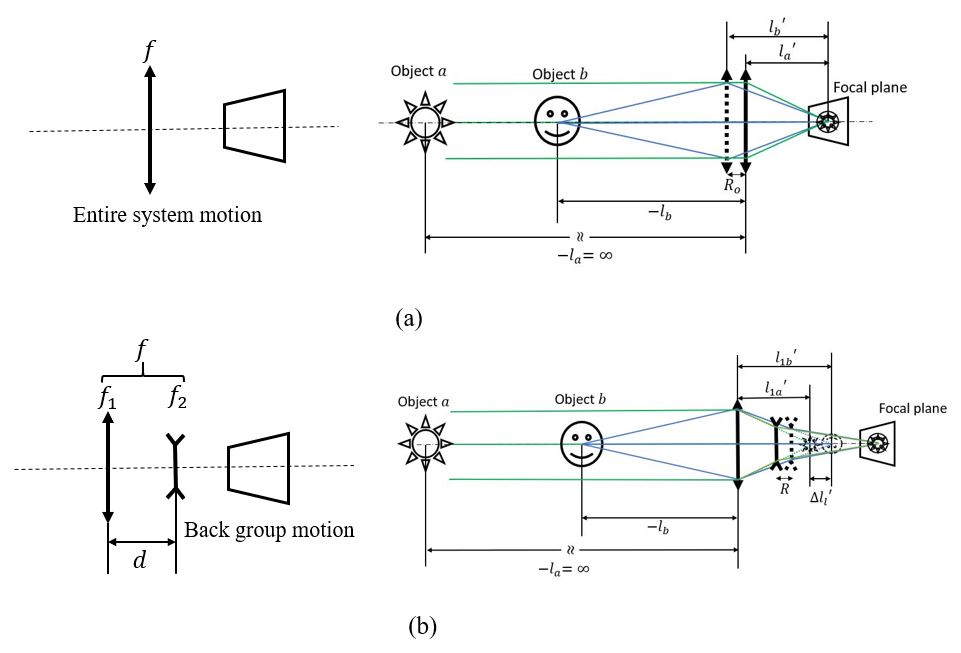}
\caption{Focusing by lens motion. (a)Focusing by translating the whole lens elements. (b)Focusing by translating only a back group lens with negative power.}
\label{fig4}
\end{figure}

As shown in Fig.\ref{fig4}, the far focusing point for majority of lenses is designated at infinity, i.e., $l_a=-\infty$, consequently $R_o=-f^2\frac{1}{l_b}$. 

Eqn.(\ref{Six_Eqn}) indicates that the traveling range is equal to the squared focal length times the diopter difference between the two positions. For example, for a lens of focal length $f=30mm$ to be focused from infinity to 2m requires a traveling range of about $450\mu m$. This traveling range is beyond the capacity of most MFMs. The solution is to avoid subjecting the whole lens assembly to motion and instead to assign the focusing task only to a sub lens group.

The simplest case is a two group lens system model, as illustrated in Fig.\ref{fig4}.(b). The lens system consists of two lens groups, the front group has no moving agent and stays stationary while the back group is driven by a actuator and assigned for focusing. The focal length of the front group is denoted as $f_1$ and that of the back group is $f_2$, the total effective focal length $f$ can be expressed as

\begin{equation}
f=\frac{f_1 f_2}{f_1+f_2-d}
\label{Sev_Eqn}
\end{equation}

Where $d$ is the separation between the two groups. 
For a practical lens system, the total effective focal length should be specified by the application and is usually positive if used with a physical image sensor. Meanwhile, the choices on $f_1$,$f_2$ and $d$ are in the discretion of the designer subjected to certain constraints. For our purpose in this work, we desire a large aperture for the front group to ensure a practical system f-number and a small aperture for the back group to fit into the small package of the MFM. Therefore, a positive lens for the front group is necessary. As for the back group, the current conditions and constraints have no preference on a positive or negative lens power yet. However, there are two other conditions which could help to narrow the design space further. First, the intermediate image from the front group should fall behind the second group for compactness. Second, the final image should be a real one to be received by an image detector. To satisfy these two conditions, for a positive back group, its object position should fall into the interval of $(0,+\infty)$, while for a negative back group, its object position is dictated to fall into the interval of $(0,-f)$. According to Eqn.(\ref{Fou_Eqn}), the expression of the image distance $l^\prime$ are the same for both cases.  

To summarize, for the front group, its condition falls into $l\in [-\infty, -2f]$ and $f>0$. While for the back group its condition falls into $l\in (0,+\infty)$ and $f>0$ or $l\in (0,-f)$ and $f<0$. After successfully narrowing the design space to a manageable size, our next task is to find solutions of reduced traveling range inside this design space. The straightforward way is to write down the expression for the traveling range as a function of parameters of the lens system. This function could reveal information on which lens parameter could be tweaked for reducing the traveling range.  

The traveling range is the difference of the image distances of the far and near object positions. The position of the final image can be derived by applying the corresponding equations in Eqn.(\ref{Fou_Eqn}) twice. The first one is to calculate the position of the intermediate image formed by the front group. Then treating this intermediate image as the object of the back group, applying the formula again on the back group, then the final image position can be acquired. In our case, the converging light of the intermediate image hits the back group before it comes to a focus. However, this does not disrupt the calculation at all since the sign convention takes care of all situations. 

Assuming our two groups lens camera changes focusing position over a range from $l_a=-\infty$ to $l_b$ and assuming $-l_b\approx L_b\gg 4|f_1|$. Under this assumption, the distance disparity of the intermediate images can be written as

\begin{equation}
\Delta {l_1}^\prime\approx-{f_1}^2\frac{1}{l_b}=(\frac{f_1}{f})^2R_o
\label{Eig_Eqn}
\end{equation}

The intermediate imaging distances of objects a and b are ${l_{1a}}^\prime=f_1$ and ${l_{1b}}^\prime={l_{1a}}^\prime+\Delta l_1^\prime$ respectively as shown in Fig.\ref{fig4}(b). To avoid confusion and abusive use of notations, let us explain our method of naming variables here. The subscripts 1 and 2 signify the quantities which are related with the first lens group and the back lens group respectively. Subscripts $a, b$ are associated the two corresponding objects. This notation method will continue to be implemented over the rest of the manuscript. 

The intermediate images are to be treated as the objects of the back group. Observing Eqn.(\ref{Ele_Eqn}) reveals that the position disparity of the intermediate images $\Delta {l_1}^\prime$ is a small quantity under the assumption $|l_b|\gg f_1$. Therefore, instead of applying Eqn.(\ref{Fou_Eqn}) directly, we employ the first order derivative approximation as shown in Eqn.(\ref{Nin_Eqn}). Not only this could save us from messy math but also benefit for revealing the major essence of the relationship among the many parameters.

\begin{equation}
\Delta {l_2}^\prime\approx\frac{\partial {l_2}^\prime}{\partial {l_1}^\prime}\Delta {l_1}^\prime
\label{Nin_Eqn}
\end{equation}

Since the final image plane is fixed in position, we have $\frac{\partial {l_2}^\prime}{\partial {l_1}^\prime}=\frac{\partial {l_2}^\prime}{\partial {L_2}}$. Combining Eqn.(\ref{Fou_Eqn}) and (\ref{Nin_Eqn}), the traveling range of focusing in our solution can be derived as

\begin{equation}
R=|\Delta {l_2}^\prime|=\frac{1}{2}\mid 1+\frac{L_2-2f_2}{\sqrt{{L_2}^2-4L_2f_2}}\mid (\frac{f_1}{f})^2R_o
\label{Ten_Eqn}
\end{equation}
Where
\begin{equation}
L_2= \frac{-(f_1-d)^2}{f_1+f_2-d}
\label{Ele_Eqn}
\end{equation}

Let $f_1=\alpha f$ and $d=\beta f$, a useful and physically viable lens system requires $f>0, d>0$, i.e., $\beta >0$. Additionally, our previously limited design space requires $\alpha>\beta$. Defining traveling range ratio as $\gamma=\frac{R}{R_o}$, then combining Eqn.(\ref{Sev_Eqn}), (\ref{Ten_Eqn}) and (\ref{Ele_Eqn}), we have

\begin{equation}
\gamma = |\frac{\alpha^2}{1-\alpha^2}|
\label{Twe_Eqn}
\end{equation}

\begin{figure}[ht!]
\centering\includegraphics[width=6cm]{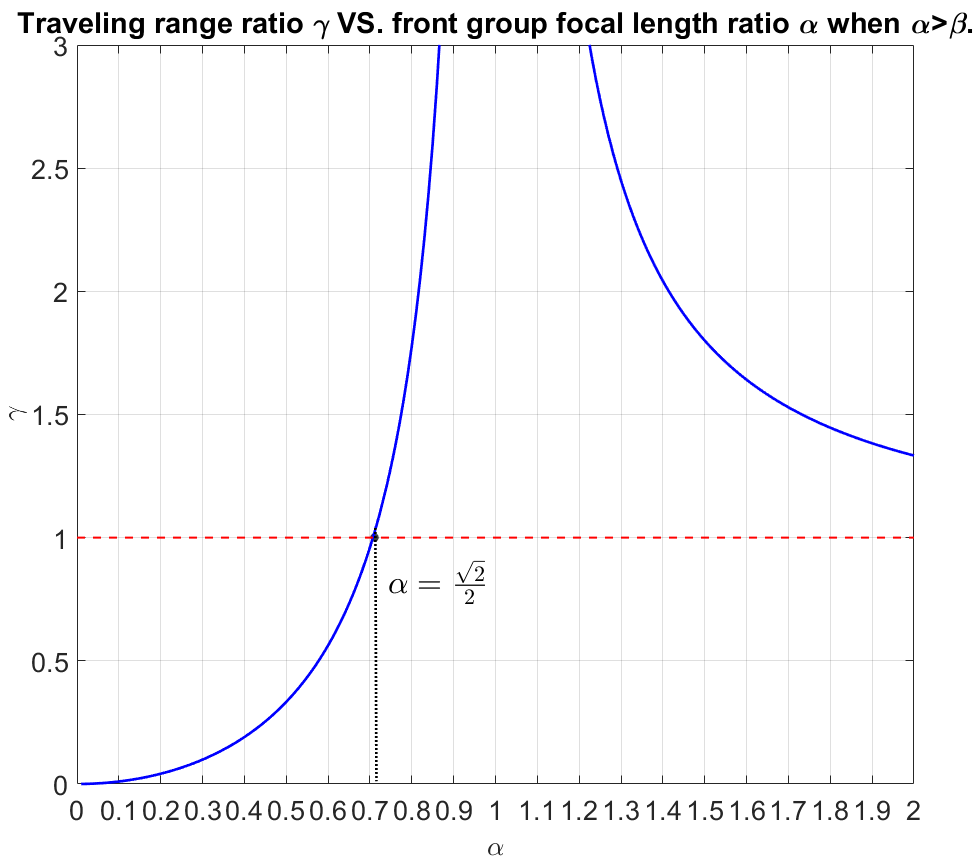}
\caption{Traveling range ratio VS. focal length ratio of front group.}
\label{fig5}
\end{figure}

Fig.\ref{fig5} shows $\gamma$ as function of $\alpha$ in our design space. $\gamma$ is monotonically increasing as $\alpha$ increases on the left half of the plot, and $\alpha=\frac{\sqrt{2}}{2}$ is the critical point where $\gamma=1$. Thus, $f_1<\frac{\sqrt{2}}{2}f$ leads to a reduced traveling range. This also dictates a back focusing group with negative power which results in a telephoto lens style. However, it is obvious that $f_1$ cannot be made arbitrarily small, since short focal length would lead to difficulty achieving targeted system aperture size. The right half of the plot in Fig.\ref{fig5} also indicates a reduced traveling range can never be achieved in case of $f_1>f$ or $f_2>0$. 

\section*{Design example}
\label{DesignExample}
\subsection{Design details}
\begin{figure}[ht!]
\centering\includegraphics[width=7cm]{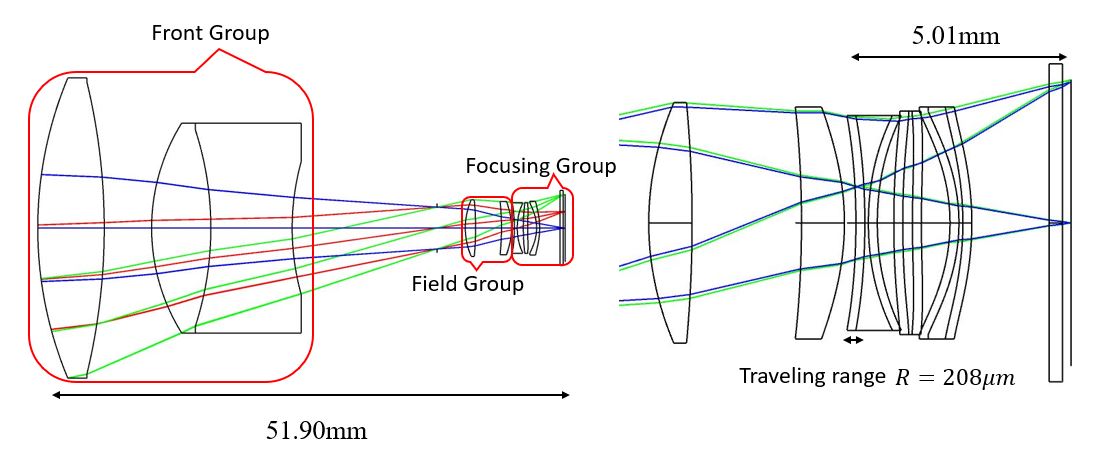}
\caption{Layout of the design example of the MFM lens with its highlighted MFM optics.}
\label{fig6}
\end{figure}

Here we present a lens design example shown in Fig.\ref{fig6}. The image sensor is of type $1/2.8$ IMX335 diagonal $6.52mm$ from SONY security product line. The effective pixel number is about $5.14M$ ($2616\times 1964$) and pixel pitch is $2\mu m \times2\mu m$. The design instance has overall focal length of $f=30mm$, $FoV=6.2^\circ$ and $F/\#=3$. The focal length of the front group is $f_1=17.49mm$ and that of the back group (focusing group) is $f_2=-5.43mm$ and separation between them is $d=15.23mm$. Thereafter, $\alpha=\frac{f_1}{f}\approx0.58, \beta=\frac{d}{f}\approx0.51$. Applying the result from Eqn.(\ref{Twe_Eqn}), the traveling range ratio $\gamma\approx0.51$.  

\begin{table}[ht!]
\centering
\caption{Prescription data for Design example of the MFM Lens.}
  \label{tab:table1}
  \begin{adjustbox}{width=\textwidth}
\begin{tabular}{c c c c c c}
\toprule
Surf.\# & Comment & Rad of Curv. $(mm)$ & Thick. $(mm)$ & Material & Diam. $(mm)$ \\ \specialrule{1pt}{1pt}{1pt}
1 & FrontGroup & 38.400  & 6.544 & FK54 & 29.568 \\
2 &  & -61.338 & 4.662 & Air & 28.824 \\ 
3 &  & 19.527 & 5.829 & FK54 & 20.687 \\
4 &  & -31.401 & 8.000 & LAF2 & 19.381 \\
5 &  & 24.485 & 14.264 & AIR & 12.987 \\
6 & IR cut Stop & Infinity & 2.775& AIR& 4.118 \\
7 & FieldGroup & 6.831& 1.010& E48R & 5.591\\
8 &  & -29.648 & 2.547 & AIR & 5.582\\
9 &  & -26.814 & 1.000 & E48R & 5.235\\
10 &  & -6.154 & 0.476 & AIR & 5.226\\
11 & FocusingGroup & -16.190 & 0.291  & POLYCARB & 4.994\\
12 &  & 6.607 & 0.593 & AIR & 4.870 \\
13 &  & 25.175 & 0.433 & E48R &  4.943\\
14 &  & 26.474 & 0.874 & AIR & 5.001\\
15 &  & -4.385 & 0.300 & POLYCARB & 5.043\\
16 &  & -7.208 & 1.802 & AIR & 5.302\\
17 & Protective Window  & Infinity & 0.307 & BK7 & 6.471\\
18 &   & Infinity & 0.201 & AIR  & 6.567\\
19 & Image & Infinity & - &    & 6.671\\ \specialrule{1pt}{1pt}{1pt} \specialrule{1pt}{1pt}{1pt}
Surf.\# & Comment & Conic & $r^4$ Aspheric\: term  & $r^6$ Aspheric\: term & $r^8$ Aspheric\: term\\
11  & FocusingGroup & 11.893 & $2.028\times10^{-4}$ & $-7.045\times10^{-5}$ & $6.732\times10^{-6}$\\
12  & Aspheric terms & 2.994 & $1.953\times10^{-3}$ & $-3.323\times10^{-4}$ & $-5.746\times 10^{-6}$\\
13  &   & 16.982 &  $1.808\times 10^{-3}$ & $-1.905\times 10^{-4}$ & $-1.245\times 10^{-5}$\\ 
14  &       & -4.776 & $-3.561\times 10^{-3}$ & $6.560\times 10^{-5}$ & $1.292\times 10^{-6}$\\
15  &       & -3.496 & $-3.961\times 10^{-3}$ & $2.635\times 10^{-4}$ & $1.624\times10^{-5}$ \\
16  &       & 0.293 & $1.417\times10^{-3}$ & $3.845\times10^{-5}$ & $1.088\times10^{-5}$\\
\bottomrule
\end{tabular}
\end{adjustbox}
\end{table}

As shown on Table\ref{tab:table1}, this design example consists of a front group, a field group and a focusing group(back group). The field group and focusing group are made out of molded polycarbonate ($n_d=1.586,V_d=29.91$) and Zeonex (Louisville, KY) E48R ($n_d=1.531, V_d=56.04$) and the elements in focusing group use even-polynomial aspheric surfaces. The front group has large aperture size and is made out of glass elements from Schott catalog. This lens can focus from $2m$ of near point to far point at infinity. Over the total focusing shift, the focusing group travels a range of $209\mu m$ corresponding to $\gamma=0.46$ which is slightly less than the calculated value $0.51$. This small difference can be accounted for by the depth of field (DoF) of lens which allows the focal plane to be slightly off the idea position.

\begin{figure}[ht!]
\centering\includegraphics[width=14cm]{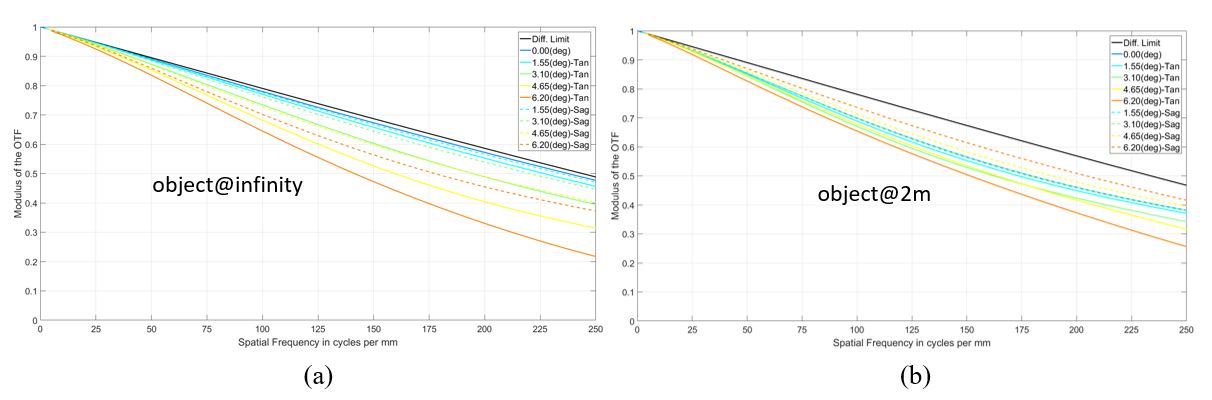}
\caption{MTF of the lens.(a)MTF of far point when focusing at infinity.(b)MTF of near point when focusing at 2m distance.}
\label{fig7}
\end{figure}

Fig.\ref{fig7} shows the modulation transfer functions (MTFs) of our lens example when object is at infinity and $2m$ respectively. The plots extend to the Nyquist frequency of the image sensor at $250$ cycles/mm.

It's worth noticing that our design resembles the accessory lens which is used to modify the focal length of  cell phone camera by attaching it to the front. These products are clipped on the cell phone camera without considering precise alignment\cite{FilterGrade}. Therefore, the MTFs deteriorate significantly thus leading to corrupted image quality. Careful tolerancing of the alignment between the front group and focusing group shows that to guarantee high imaging quality, the tilt tolerance is about $\pm 0.2^\circ$ and the decenter tolerance is about $\pm 25\mu m$. This alignment tolerance is achievable in mass-production but not likely by clipping accessory lenses.       

\subsection*{Universal focusing}
Conventionally, elements from same lens are highly correlated and inseparable since they are meant to modulate the light beam and correct aberrations collectively as intended during the design stage. It is rare that lens elements can be disassembled and reused in another different lens design, especially when high MTFs are required. However, in our design scheme, the light has been preprocessed and shaped substantially before it enters the focal group as shown in Fig.\ref{fig4}. This preprocessing has the potential for preparing a standard light of beam for a universal back group processing.  

Universal focusing can reduce the development cost significantly. The optics in miniature cameras is commonly composed of aspheric plastic elements made out of injection molding\cite{neefe1981injection}. Making molds accounts for a large proportion of cost. Universal focusing helps cut off the number of molds needed therefore cutting off that portion of the cost.

As discussed in previous sections, implementing MFMs imposes rigid constraints and limitations on lens specifications. A family of lenses employing a common MFM share some constraints. First of all, the traveling range enabled by the actuator is fixed, therefore, the scene depths of each lens should be subjected to a same traveling range limit. Secondly, the highest f-stop or aperture denoted by $F/\#$ is limited by this common MFM. Lastly, since image sensor is built into the miniature focusing module, the marginal $FoV$ of each lens will be subjected to same maximum image height.

\begin{figure}[ht!]
\centering\includegraphics[width=10cm]{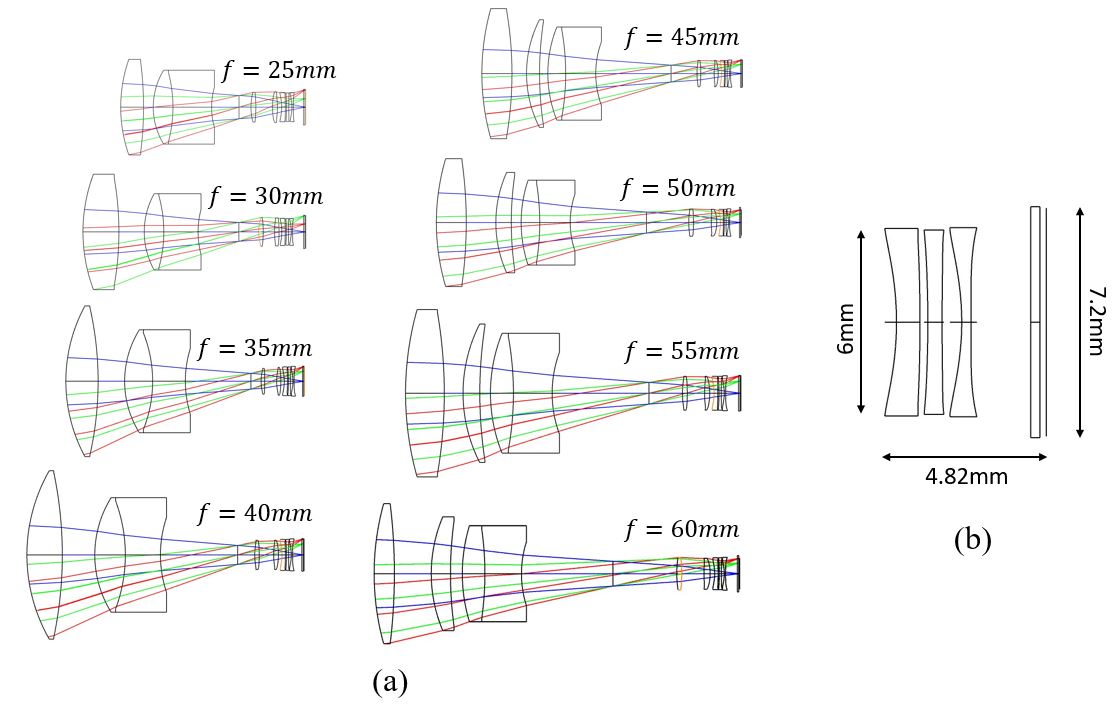}
\caption{Universal focusing.(a)Different lens systems using a common miniature focusing group.(b)Highlighted common miniature focusing group.}
\label{fig8}
\end{figure}

 Fig.\ref{fig8}(a) presents a collection of eight lenses are presented with focal length ranging from $25mm$ to $60mm$. The focusing group of each lens is implemented using an identical optics design as highlighted in Fig.\ref{fig8}(b). The dimension of this focusing group is optimized for fitting into a typical VCM encapsulation. The focusing parameters are provided on Table.\ref{tab:table2}. The last column on Table.\ref{tab:table2} shows all $\gamma<1$ which implies a reduced traveling range.

\begin{table}[ht!]
\centering
\caption{Characteristics of universal focusing on a group of lenses of varying specs.}
  \label{tab:table2}
\begin{tabular}{c c c c c c c}
\toprule
Lens.$\#$ & f(mm) & $F/\#$ & $-l_b(m)$ & $R(\mu m)$ & $R_o(\mu m)$ & $\gamma$\\
1 & 25 & 2.5 & 2 & 238 & 313 & 0.76\\
2 & 30 & 3& 3 & 207 & 300 & 0.69\\
3 & 35 & 3 & 4 & 238 & 306 & 0.78\\
4 & 40 & 3 & 5 & 252 & 320 & 0.79\\
5 & 45 & 3.5 & 6 & 224 & 338 & 0.66\\
6 & 50 & 3.5 & 8 & 204 & 313 & 0.65\\ 
7 & 55 & 3.5 & 10 & 208 & 303 & 0.69\\
8 & 60 & 4 & 12 & 205 & 300 & 0.68\\
\bottomrule
\end{tabular}
\end{table}

\section*{Digital zoom with array cameras}
\subsection{Two basic configurations of array cameras}
The implementation of MFM imposes strict limitation on the size of image sensor. Therefore, such a lens can only capture relatively narrow $FoV$. Moreover, it is too much challenging to perform optical zoom with small lenses. Instead of using a single lens, we build parallel cameras with arrays of lenses. Parallel cameras synthesize wide field image by stitching image units from multiple channels and provides a panoramic view imitating the function of 'zoom-out'. Meanwhile, with small pixel pitch and lens of high MTF performance, each camera generates high angular resolution image which supports a deep digital zoom-in capacity. As described in \cite{brady2018parallel}, parallel cameras present novel opportunities by radically increasing the information capacity of cameras in terms of pixel count per unit volume, weight and cost. In practice, large portion of cost of cameras goes to the focus and zoom mechanisms. Through parallel architecture, cost associated with optical zoom is largely eliminated and optical zoom can be replaced by more efficient digital zoom.

\begin{figure}[ht!]
\centering\includegraphics[width=14cm]{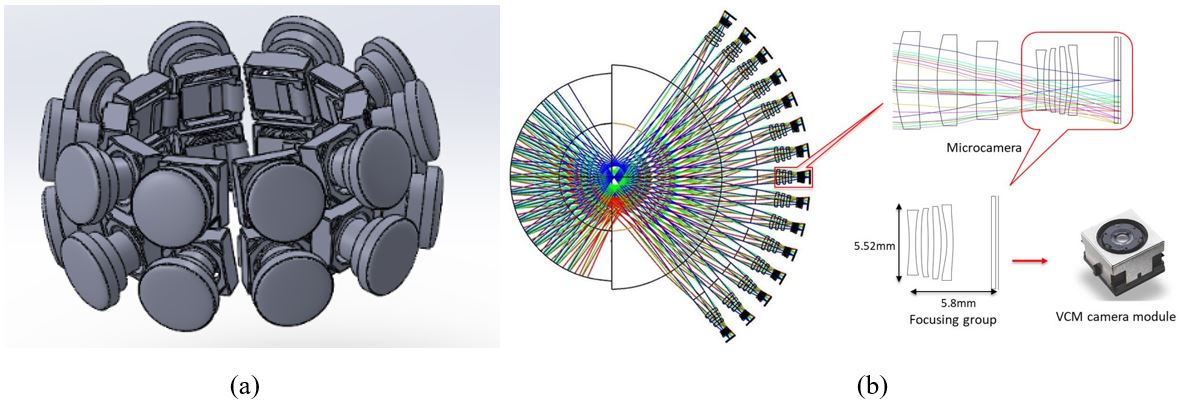}
\caption{Parallel cameras.(a)Cameras array with MFM.(b)Monocentric multiscale lens with MFM.}
\label{fig9}
\end{figure}

Parallel cameras consist of numerous camera units. Each unit is acting like a building block. This arrangement offers high flexibility on building cameras of versatile specifications and ability of easy re-configuration\cite{pang2018field}. As shown in Fig.\ref{fig9}(a), optically independent lenses are positioned over a pre-designed angular grid with each one covering a corresponding angular subsection. In Fig.\ref{fig9}(b), we illustrate a monocentric multiscale (MMS) architecture\cite{brady2009multiscale}. MMS cameras consist of two parts, one is monocentric objective lens with large aperture in the front and the other is multiple microcamera array in the back. Since the objective lens is shared by all the array units, this architecture is more compact and efficient in context of information per volume. 

\subsection*{A design example of a MMS lens with high digital zoom ratio}
An inexpensive and simple focusing solution for MMS lens architectures is highly desireable. Several previous trials relying on piezoelectric motor have been successful but too expensive for mass production \cite{brady2012multiscale,marks2014characterization,llull2015characterization}. MFM based on VCM actuators provide a promising chance resolve this challenge. Here we present a  design example to demonstrate the effectiveness of this new approach which enables a low cost parallel camera capable of high digital zoom ratio. 

\begin{figure}[ht!]
\centering\includegraphics[width=9cm]{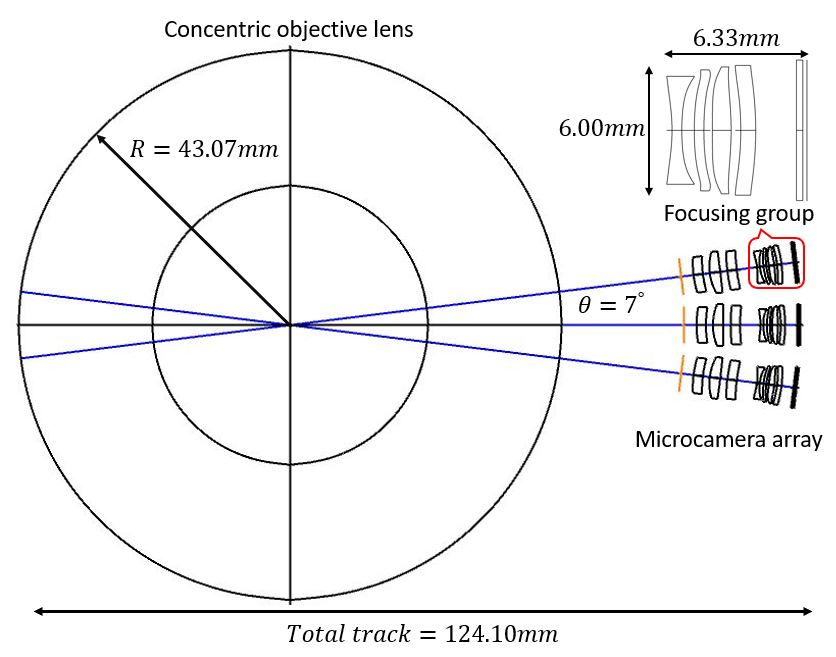}
\caption{Layout and critical dimensions of the design example of MMS lens with built in MFM.}
\label{fig10}
\end{figure}
\begin{table}[hbt!]
\centering
\caption{Prescription data for Design example of the MMS Lens with built in MFM.}
  \label{tab:table3}
\begin{adjustbox}{width=\textwidth}
\begin{tabular}{c c c c c c}
\toprule
Surf.\# & Comment & Rad of Curv. $(mm)$ & Thick. $(mm)$ & Material & Diam. $(mm)$ \\ \specialrule{1pt}{1pt}{1pt}
1 & Concentric objective lens & 43.074 & 21.214 & S-NBH8 & 86.000 \\
2 &  & 21.860 & 21.860 & F\_SILICA & 43.720 \\ 
3 &  & Infinity & 21.860 & F\_SILICA & 43.720 \\
4 &  & -21.860 & 21.214 & S-NBH8 & 43.720 \\
5 &  & -43.074 & 19.495 & AIR & 86.000 \\
6 & IR cut Stop & Infinity & 1.970& AIR& 5.802 \\
7 & FieldGroup & 15.950 & 1.487 & N-SK16 & 5.600\\
8 &  & 17.695 & 1.107 & AIR & 5.600\\
9 &  & 8.333 & 1.677 & N-FK51A & 6.535\\
10 &  & -177.672 & 1.145 & AIR & 6.395\\
11 &  & 27.868 & 1.498 & N-LASF40 & 6.087\\
12 &  & 29.474 & 3.236 & AIR & 5.755 \\
13 & FocusingGroup  & -8.295 & 0.459 & POLYCARB &  5.046\\
14 &  & 11.633 & 0.586 & AIR & 5.017\\
15 &  & 14.297 & 0.450 & POLYCARB & 5.346\\
16 &   & 14.248 & 0.399 & AIR & 5.656\\
17 &  & 1799.46 & 0.879 & E48R & 5.931\\
18 &  & -13.138 & 0.453 & AIR  & 5.956\\
19 &  &-14.192 & 0.700 & POLYCARB & 6.066\\
20 &  & -15.853 & 1.911 & AIR & 6.520\\
21 & Protective Window & Infinity & 0.300 & N-BAK7 & 6.562\\
22 & & Infinity & 0.200 & AIR & 6.562\\
23 & Image & Infinity & - &    &  6.671\\ \specialrule{1pt}{1pt}{1pt} 
\end{tabular}
 \end{adjustbox}
\begin{adjustbox}{width=\textwidth}
\begin{tabular}{c c c c c c c}
\toprule
Surf.$\#$  & Conic & $r^4$ Aspheric\: term  & $r^6$ Aspheric\: term & $r^8$ Aspheric\: term
&  $r^{10}$ Aspheric\: term &  $r^{12}$ Aspheric\: term \\
13   & -4.764 & $2.388 \times10^{-3}$ & $2.209\times10^{-4}$ & $4.193\times10^{-6}$ & $-3.882\times10^{-6}$ & $-2.185\times10^{-7}$\\
14  & 17.002 & $4.750\times10^{-3}$ & $2.877\times10^{-4}$ & $4.107\times 10^{-6}$ & $2.009\times 10^{-6}$ & $-1.306\times 10^{-6}$\\
15  & 16.557 &  $1.739\times 10^{-3}$ & $1.637\times 10^{-4}$ & $-1.663\times 10^{-5}$ & $3.536\times 10^{-6}$ & $-1.374\times 10^{-6}$ \\ 
16  & -7.438 & $1.562\times 10^{-3}$ & $1.411\times 10^{-4}$ & $1.729\times 10^{-5}$ & $1.056\times 10^{-5}$ & $-2.318\times 10^{-6}$ \\
17   & -20.000 & $3.987\times 10^{-3}$ & $2.139\times 10^{-4}$ & $-4.189\times 10^{-5}$ & $1.361\times 10^{-5}$ & $-1.080\times 10^{-6}$ \\
18  & 15.412 & $4.494\times 10^{-3}$ & $-1.413\times 10^{-4}$ & $-2.286\times 10^{-6}$ & $3.025\times 10^{-6}$ & $-6.047\times 10^{-8}$\\
19   & 16.652 & $6.235\times 10^{-4}$ & $-2.146\times 10^{-5}$ & $1.025\times 10^{-5}$ & $4.113\times 10^{-6}$ & $-3.275\times 10^{-7}$\\
20    & 20.000 & $9.242\times10^{-4}$ & $1.998\times10^{-4}$ & $-3.570\times10^{-5}$ & $1.621\times10^{-6}$ & $1.194\times10^{-7}$\\
\bottomrule
\end{tabular}
 \end{adjustbox}
\end{table}

The design work starts out with the same image sensor 1/2.8 SONY IMX335 used in our previous lens example in Sec.\ref{DesignExample}. We choose an overall focal length of $f=45mm$, aperture size of $F/\#=3$ and field view of each channel of $MFoV=8.28^\circ$. The physical angle subtended by each microcamera is $\theta=7^\circ$, therefore the overlap between adjacent channels is $1.28^\circ$. The lens focuses from infinity of its far point to 5m of its near point by translating the focusing group in axial direction with traveling range of $260\mu m$ which indicates a traveling range reduction ratio of $\gamma=0.64$. Table\ref{tab:table3} shows the detailed prescription data of this design work and Fig.\ref{fig10} shows the layout of the lens system. The whole system consists of a concentric objective lens in the front and a microcamera lens array in the back adding up to a total track length of $124.10mm$ . The front objective lens is made up of two concentric spherical layers. The inner layer is a solid ball lens made of fused silica and the outer shell is made of S-NBH8 glass from Ohara catalog. The stop surface sitting between the objective lens and microcamera lens serves as an important light controller balancing the two parts of the system. The microcamera lens immediately following the stop has a stationary front group which condenses the light beams and compresses them into the focusing group with a relatively small aperture size. There are four elements in the focusing group with each one being made of aspheric plastics including polycarbonate and E48R from Zeonex. The aperture diameter of each element is controlled to be less than $6mm$ complying with the size constraint of the MFM packages.

\begin{figure}[ht!]
\centering\includegraphics[width=14cm]{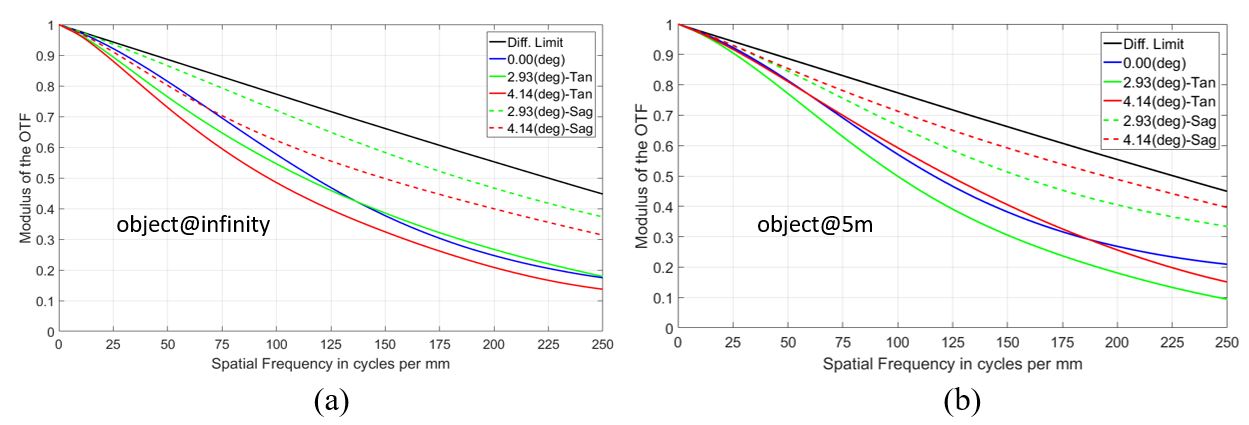}
\caption{MTF curves of the design example at two working distances.(a)MTF curves with object being at infinity.(b)MTF curves with object being at 5m.}
\label{fig11}
\end{figure}

As discussed in several previous MMS studies, the spherically symmetrical layout renders off-axis aberrations insignificant and the only aberrations of concern are spherical aberration and sphero-chromatic aberration. The aspheric profile employed in focusing group can effectively correct for these two remaining aberrations. The pixel size is $2\mu m$ and the Nyquist frequency is $250lp/mm$. As illustrated in Fig.\ref{fig11}, the contrast reproduction is above $10\%$ at the maximum frequency and above $40\%$ at $125$ lp/mm which is half of the maximum for both working distances at far point of infinity and near point of $5m$.  

Alignment and assembly of the microcamera array are issues of paramount importance. The microcamera array is packed on a spherical doom with a cone shaped space cell reserved for each imaging unit. The angle of the cone with vertex in the center of spherical objective lens measures the space allotted to each unit, which is called physical cone angle. On one hand, the physical cone angle should be large enough for fitting into the assemble of the microcamera unit and to prevent physical conflict between neighboring units. On the other hand, it should be controlled for preventing image gaps and ensuring overlap between neighboring microcamera field of view (MFoV). As illustrated in Fig.\ref{fig10}, the physical cone angle in our design work is $\theta=7^\circ$ which implies a overlap of $1.28^\circ$. The first surface of the focusing group is about $75mm$ away from the vertex which provides a lateral dimension of $9.17mm$ inside the cone. This is sufficient for accommodating a typical VCM assembly. 

Tight tolerance requirement on focusing group alignment increases cost and thus hurt productivity. Fortunately, in our design example, the decenter is about $\pm 30\mu m$ and the tilt is about $0.25\circ$ which are commercial precision and can be easily satisfied in mass production. 

The MMS lens can easily achieve a total FoV coverage greater than $100^\circ$ by piecing together a large amount of microcameras. In our design example, the instantaneous FoV is about $45\mu rad$ for each pixel which indicates an equivalent focal length of $180mm$ for a full frame camera. To summarize, with a imaging system being capable of wide angle and telephoto simultaneously, a superior digital zoom capacity is easily guaranteed.

\section*{Conclusion}
Focus and zoom mechanisms comprise significant part of the total cost of traditional lenses. Over the past 20 years, the technology utilized in traditional cameras has improved much more slowly than that of miniaturized cameras. By borrowing focusing mechanism from miniaturized camera, the size and cost of cameras tends to be reduced considerably. This paper discusses the challenges along with transferring miniaturized technology and possible solutions to handle these challenges. Small sensor format and the implementation of field lens help produce valid lens solution under highly constrained small space. By balancing back group focusing and focal power distribution, a reduced traveling range is achievable. And finally, With parallel camera architecture, a new paradigm of camera design enables simultaneous wide $FoV$ and high angular resolution image throughput which indicates greatly increased space-bandwidth product. This high information throughput allows deep range digital zoom without compromising image quality. 

However, small format sensor and highly constrained space, which are unavoidable with MFM, impose a limitation on overall focal length of the system. As focal length increases, useful aperture size and $FoV$ become hard to achieve. Even with well corrected aberration, the diffraction effect from large $F/\#$ would flood the small pixel pitch. In addition, the $FoV$ will become too narrow and alignment of different units in parallel camera becomes too much challenging.   

%%%%%%%%%%%%%%%%%%%%%%% References %%%%%%%%%%%%%%%%%%%%%%%%%
%%%%%%%%%% If using BibTeX:
\bibliography{Focus.bib}
\bibliographystyle{ieeetr}
%%%%%%%%%% If preparing manually:
% \newcommand{\enquote}[1]{``#1''}

% \bibitem{Zhang:14}
% Y.~Zhang, S.~Qiao, L.~Sun, Q.~W. Shi, W.~Huang, L.~Li, and Z.~Yang,
%   \enquote{Photoinduced active terahertz metamaterials with nanostructured
%   vanadium dioxide film deposited by sol-gel method,}
%   {\protect\JournalTitle{Optics Express}} \textbf{22}, 11070--11078 (2014).

% {Optical Society}, \enquote{{OSA Publishing},}
%   \url{http://www.osapublishing.org}.

% \bibitem{FORSTER2007}
% P.~Forster, V.~Ramaswamy, P.~Artaxo, T.~Bernsten, R.~Betts, D.~Fahey,
%   J.~Haywood, J.~Lean, D.~Lowe, G.~Myhre, J.~Nganga, R.~Prinn, G.~Raga,
%   M.~Schulz, and R.~V. Dorland, \enquote{Changes in atmospheric consituents and
%   in radiative forcing,} in \enquote{Climate Change 2007: The Physical Science
%   Basis. Contribution of Working Group 1 to the Fourth assesment report of
%   Intergovernmental Panel on Climate Change,}  S.~Solomon, D.~Qin, M.~Manning,
%   Z.~Chen, M.~Marquis, K.~B. Averyt, M.~Tignor, and H.~L. Miler, eds.
%   (Cambridge University Press, 2007).

% \end{thebibliography}
\end{document}